\documentclass{article}
\usepackage[T1]{fontenc} 
\usepackage[utf8]{inputenc} 
\usepackage{ismir,amsmath,cite,url}
\usepackage{url}
\usepackage{graphicx}
\usepackage{color}
\usepackage{array}
\usepackage{multirow}
\usepackage{lineno}
\usepackage{subfig}
\usepackage{amssymb}
\usepackage{amsmath}

\title{Timbre Transfer using Image-to-Image Denoising Diffusion Implicit Models}

\multauthor
{Luca Comanducci \hspace{1cm} Fabio Antonacci \hspace{1cm} Augusto Sarti} {
 Dipartimento di Elettronica, Informazione e Bioingegneria, Politecnico di Milano, Italy\\
{\tt\small luca.comanducci@polimi.it, fabio.antonacci@polimi.it, augusto.sarti@polimi.it}
}

\def\authorname{L. Comanducci, F. Antonacci, and A. Sarti}

\usepackage[bookmarks=false,pdfauthor={\authorname},pdfsubject={\papersubject},hidelinks]{hyperref}

\begin{document}

\maketitle
\begin{abstract}
Timbre transfer techniques aim at converting the sound of a musical piece generated by one instrument into the same one as if it was played by another instrument, while maintaining as much as possible the content in terms of musical characteristics such as melody and dynamics. Following their recent breakthroughs in deep learning-based generation, we apply Denoising Diffusion Models (DDMs) to perform timbre transfer. Specifically, we apply the recently proposed Denoising Diffusion Implicit Models (DDIMs) that enable to accelerate the sampling procedure. 
Inspired by the recent application of DDMs to image translation problems we formulate the timbre transfer task similarly, by first converting the audio tracks into log mel spectrograms and by conditioning the generation of the desired timbre spectrogram through the input timbre spectrogram.  
We perform both one-to-one and many-to-many timbre transfer, by converting audio waveforms containing only single instruments and multiple instruments, respectively.
We compare the proposed technique with existing state-of-the-art methods both through listening tests and objective measures in order to demonstrate the effectiveness of the proposed model.

\section{Introduction}
\begin{figure*}
    \centering
    \includegraphics[width=\textwidth]{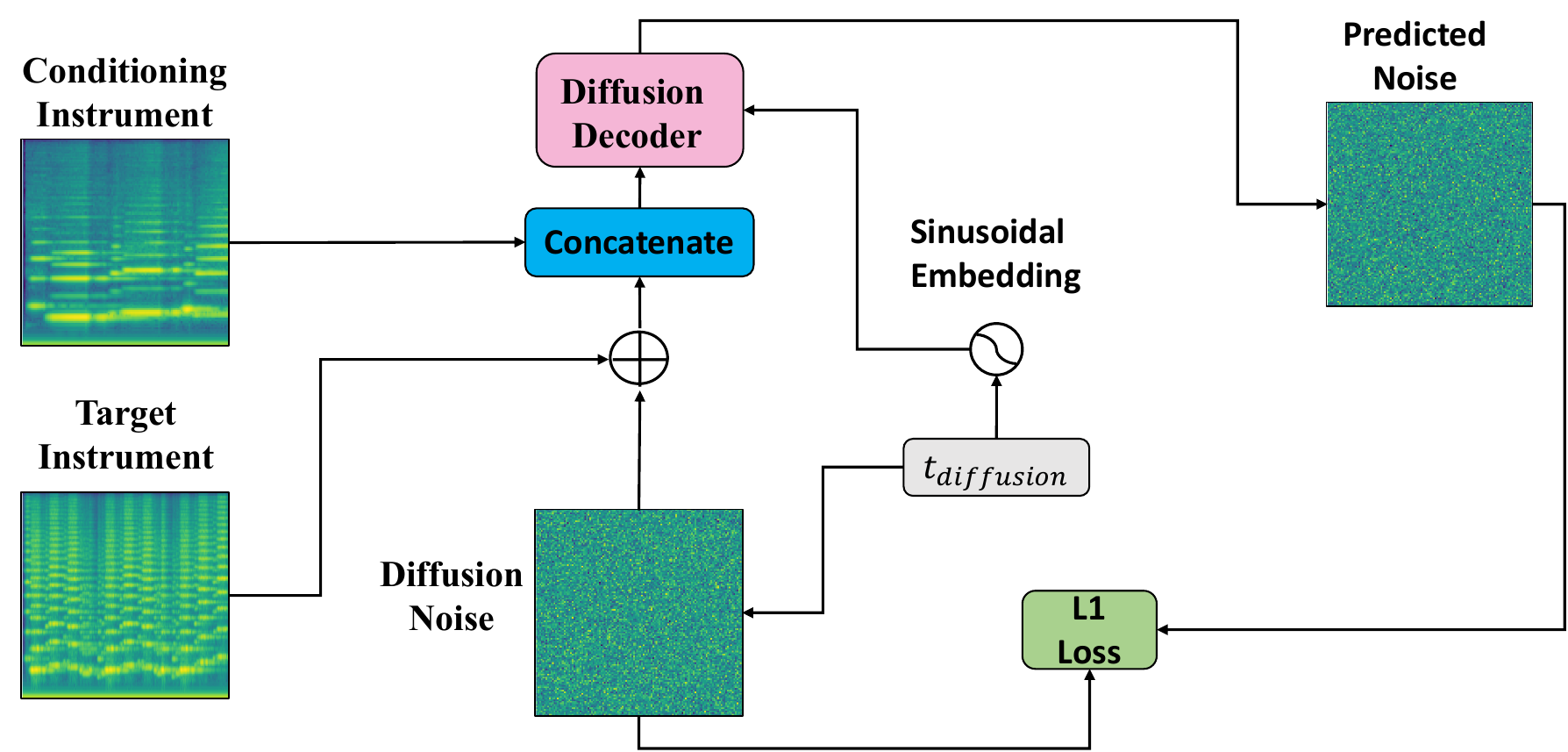}
    \caption{Training scheme of the proposed DiffTransfer technique. The target instrument spectrogram is summed with noise following a simplified cosine schedule. The decoder, conditioned on the conditioning instrument spectrogram and on the sinusoidal embedding representing the current time instant estimates the added noise. The decoder parameters are estimated by computing the L1 loss between the ground truth and the estimated diffusion noise. }
    \label{fig:training_scheme}
\end{figure*}
Timbre is an extremely important perceptual aspect of music, yet it is hard to both model and define. 
The concept of musical timbre can be defined as the perceived characteristics of a musical sound that are different from pitch and amplitude contours~\cite{colonel2020conditioning}.

Timbre Transfer concerns the task of converting a musical piece from one timbre to another while preserving the other music-related characteristics. While this operation is not trivial, it is of extreme interest for several applications, from the development of plugins to be used in Digital Audio Workstations (DAW) to enabling the possibility of playing sounds corresponding to not widely available musical instruments. 

In this paper, we present DiffTransfer, a technique for timbre transfer which is tested both between single and multiple instruments and is based on a continuous Denoising Diffusion Implicit Model (DDIM) with deterministic sampling~\cite{song2021denoising}, a modified version of Denoising Diffusion Probabilistic Models (DDPMs) that are trained using the same procedure, but allow for faster sampling times. Specifically, in~\cite{song2021denoising} it was empirically shown that DDIMs allow for $10\times-50\times$ faster wall-clock time performances with respect to DDPMs.

In order to be able to convert one timbre into another, we use a procedure similar to the recently proposed image-to-image technique Palette~\cite{saharia2022palette}. Specifically, we use as input to the diffusion model the noise and condition it with the chosen input timbre spectrogram, then, through the denoising procedure, the model learns to reconstruct spectrograms of the desired timbre. We consider the scenario where the timbre-transfer task is \textit{paired}, which means that the desired and input spectrograms have the same melodic/harmonic content, but differ in terms of timbre.

We experiment both with the possibility of converting between tracks containing only single instruments and also mixtures of instruments, with no prior separation step, while making no modifications to the model in order to take into account both configurations. 

In order to demonstrate the effectiveness of the proposed model, we compare DiffTransfer with state-of-the-art techniques, both through objective measures and by performing a user-based listening test.
The source code and audio excerpts can be found at \url{ https://lucacoma.github.io/DiffTransfer/}.
\end{abstract}
\section{Related Work}
Several types of timbre Transfer techniques have been proposed in the literature.
In~\cite{huang2018timbretron} a CycleGAN~\cite{zhu2017unpaired} is applied in order to perform an unpaired transfer using the Constant-Q transform and the audio is then recovered through a WaveNet~\cite{oord2016wavenet} model. In~\cite{jain2020att} an attention-based architecture is applied in order to convert mel spectrograms, which are then inverted through a MelGAN architecture~\cite{kumar2019melgan}. Gaussian mixture-based variational autoencoders are applied~\cite{luo2019learning} in order to learn a latent space where pitch and timbre representations are disentangled.

Another class of methods, instead, extracts musical parameters such as pitch and loudness from the input audio tracks and performs the transfer by resynthesizing sound through a network that has learned to generate tracks with the desired timbre. The most known example of these techniques is the Differentiable Digital Signal Processing (DDSP)~\cite{Engel2020DDSP} model. Other similar techniques were proposed such as~\cite{michelashvili2020timbre}, where a hierarchical model is used in order to reconstruct the signal at increasing resolutions.
Recently there have been proposed also models that directly work on the audio waveform such as~\cite{musictranslation}, where music pieces are translated to specific timbre domains. The only model that, to the best of our knowledge and except for the one proposed in this paper, is tested on multi-instrument timbre transfer without any source separation pre-processing is the Music-STAR network, presented in~\cite{alinoori2022musicstar}. In Music-STAR a WaveNet autoencoder~\cite{engel2017neural} is trained by applying teacher-forcing~\cite{williams1989learning} to the decoders in order to recover the desired timbre.

Denoising Diffusion Probabilistic Models (DDPMs) \cite{ho2020denoising} have recently become the latest state-of-the-art for what concerns deep learning-based generation fastly replacing Generative Adversarial Networks (GANs)~\cite{goodfellow2020generative} and Variational Autoencoders~\cite{roberts2017hierarchical}, due to their easier training procedure and increased quality of the produced results.

DDPMs have been successfully applied to a wide variety of image-related tasks such as generation~\cite{saharia2022photorealistic} and translation~\cite{saharia2022palette}.

More recently, DDPMs have been also used for audio-related tasks. In~\cite{hawthorne2022multiinstrument} a diffusion model is applied in order to convert midi tracks to spectrograms, while in~\cite{yang2022diffsound} a text-to-music diffusion model is proposed. DDPMs have also been applied to symbolic music generation~\cite{mittal2021symbolicdiffusion}, speech synthesis~\cite{kong2021diffwave} and singing voice extraction~\cite{PlajaRoglans2022}.

While DDPMs have extremely powerful generation capabilities they suffer from slow sampling times. To ameliorate this issue, recently Denoising Diffusion Implicit Models (DDIMs)~\cite{song2021denoising}, which allow for faster sampling times and were recently applied to image inpainting~\cite{zhang2023towards}.

\section{Proposed Model}
In this section, we describe the proposed DiffTransfer technique for timbre transfer. Instead of working directly with raw audio signals, we convert them into log mel-scaled spectrograms, due to their easier handling by deep learning models. We then propose a model that, given as input the spectrogram corresponding to the conditioning instrument, generates the corresponding target spectrogram that would have been obtained by playing the same piece of music with the target instrument. 
Operatively we achieve this through a conditional continuous-time DDIM, which learns to denoise the target instrument spectrogram, while conditioned on the input instrument spectrogram, as depicted in Fig.~\ref{fig:training_scheme}. At inference time, the model is fed with the input conditioning instrument concatenated with Gaussian noise and generates the corresponding target spectrogram. We retrieve the audio signal by applying to the log mel spectrograms the SoundStream\footnote{\url{https://tfhub.dev/google/soundstream/mel/decoder/music/1}} model~\cite{zeghidour2021soundstream}, provided by~\cite{hawthorne2022multiinstrument} where it was trained on a custom music dataset.

In the following, we'll provide a brief overview of the DDIM framework and notation used in this paper, in order to make the tractation as compact as possible, for additional and more thorough formulations, we refer the reader to~\cite{song2021denoising} and~\cite{saharia2022palette}. We aim at giving a general overview of the process and we'll use a slight abuse of notation to describe the diffusion process using the continuous time framework, in order to make it more similar to the more common literature regarding DDPMs and DDIMs.

\subsection{Diffusion Decoder}
\begin{figure*}[t]
    \centering
    \includegraphics[width=\textwidth]{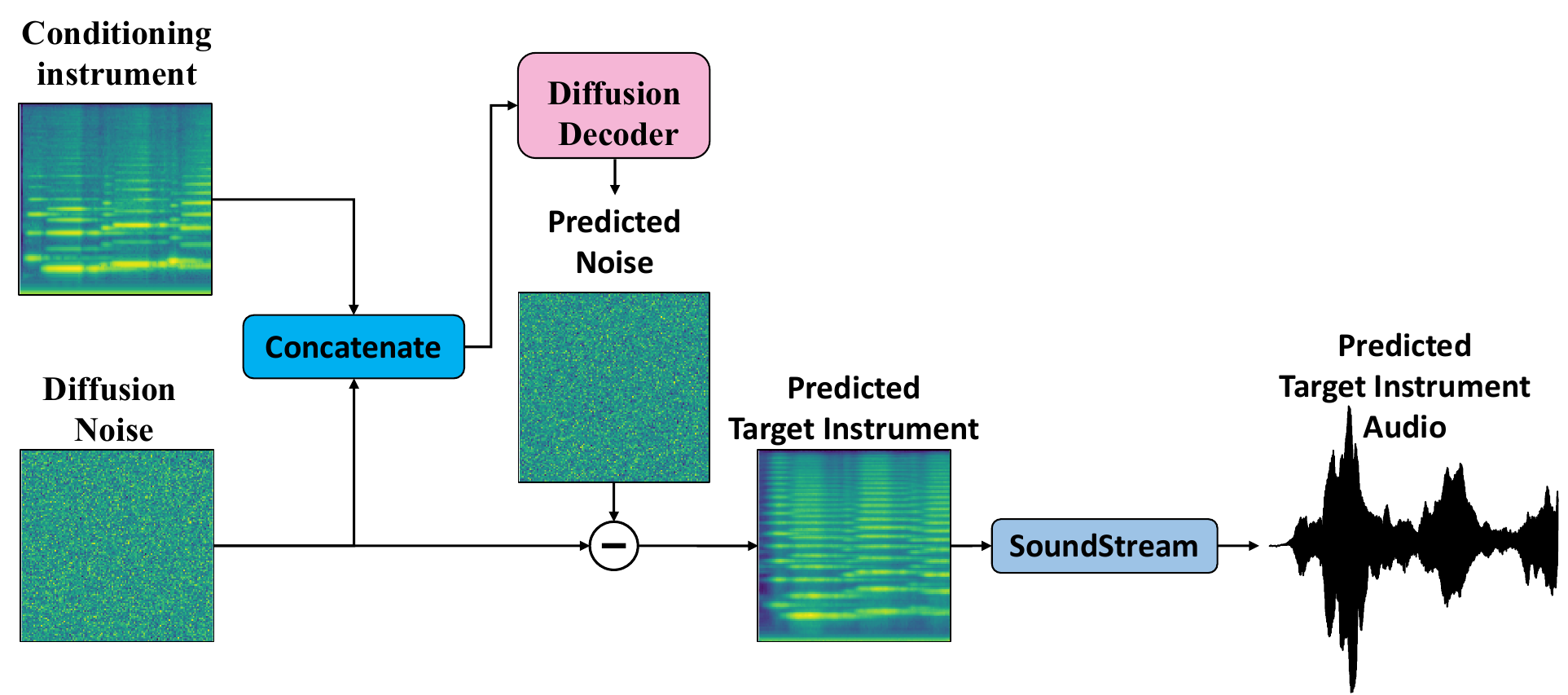}
    \caption{Deployment scheme of the proposed DiffTransfer technique. The decoder is fed with Gaussian noise and with the conditioning instrument spectrogram. The noise estimate provided by the decoder is then subtracted from the input noise in order to provide an estimate of the desired target spectrogram, from which the audio is estimated via the SoundStream model~\cite{zeghidour2021soundstream,hawthorne2022multiinstrument}.}
    \label{fig:test_scheme}
\end{figure*}
We adopt a procedure similar to the Palette~\cite{saharia2022palette} image-to-image translation technique in order to train the timbre transfer decoder as a Denoising Diffusion Implicit Model (DDIM)~\cite{song2021denoising}. Broadly speaking, DDIMs work by learning how to generate data from noise in a two-part procedure. The first part is denoted as the \textit{forward process}, where Gaussian noise $\gamma \sim \mathcal{N}(0,1)$ is subsequently added to the input until it is indistinguishable from the former. The second part consists of the \textit{reverse process} where a decoder learns how to invert the forward process, effectively reconstructing data from the noise. DDIMs can be seen as a generalization of DDPMs that shares the same training procedure, however, they differ in the modeling of the reverse process, by using a non-markovian diffusion process, which allows for faster generation times.

\subsubsection{Forward Process}

Let us define $\mathbf{X}$ and $\mathbf{Y}$ as the log mel spectrograms corresponding to the conditioning and target instruments, respectively. We choose a continuous diffusion time~\cite{campbell2022continuous,song2021scorebased,rouard2021crash}in order to be able to change the number of desired sampling steps. If we consider $T$ steps, then the diffusion time can be defined as $t \in\{0,1\}$, where consecutive times are separated by $\Delta_{t}=1/T$.
Then, the forward process is defined similarly to the case of DDPMs by subsequently adding noise to the target spectrogram for $T$ steps
\begin{equation} \label{eq1}
\begin{split}
&q(\mathbf{Y}_{t}|\mathbf{Y}_{t-\Delta_{t}}) = \mathcal{N}(\mathbf{Y}_{t}, \sqrt{(\alpha_t)}\mathbf{Y}_{t-\Delta_{t}}, \beta_t \mathbf{I}), \\
& ~~~~~~~~~q(\mathbf{Y}_{1:T}|\mathbf{Y}_{0}) = \prod_{t=1}^{T}q(\mathbf{Y}_{t-\Delta_{t}})
\end{split}
\end{equation}
where $\alpha$ and $\beta$ are parameters defined by a simplified cosine schedule~\cite{nichol2021improved}. 
\subsubsection{Reverse Process}
In the case of DDIMs, the reverse diffusion process is operated by introducing an additional distribution $p_\theta$, where a sample $\mathbf{Y}_{t-\Delta t}$ can be generated from a sample $\mathbf{Y}_t$  as
\begin{equation}
    \begin{split}
     \mathbf{Y}_{t-\Delta t} = &\sqrt{\beta_{t-\Delta t}} \left( \frac{ c-\sqrt{\beta_t}\gamma_\theta^{(t)}(\mathbf{Y}_{t},\mathbf{X})}{\sqrt(\alpha_t)}\right) + \\
     &\sqrt{1-\alpha_{t-\Delta_t}} \cdot \gamma_\theta^{(t)}(\mathbf{Y}_{t},\mathbf{X}),
     \end{split}
\end{equation},
where $\gamma$ is the noise estimated by a network with parameters $\theta$. The noise at time $t$ $\gamma_\theta^{(t)}$ is estimated by a network that is conditioned also on the input timbre spectrogram $\mathbf{X}$, similarly to the formulation proposed in Palette~\cite{saharia2022palette}.

\subsubsection{Training Procedure}
The denoising process is operated through a U-Net architecture which is conditioned on $\mathbf{X}$ and trained to predict the added noise in order to minimize the L1 loss
\begin{equation}
    \mathbb{E} = ||\gamma_\theta^{(t)}(\mathbf{Y}_{t},\mathbf{X})-\gamma||_{1}^{1},
\end{equation}
where $\gamma$ is the true perturbation, while $\gamma_\theta^{(t)}(\mathbf{Y}_{t},\mathbf{X})$ is the estimate of the noise added to the target spectrogram at time $t$, conditioned on the input spectrogram $\mathbf{X}$.

\subsection{Architecture}
The decoder architecture is based on a U-Net model. The building element is made of residual blocks, in each of these the input is processed by (i) a 2D convolutional layer with swish activation, followed by batch normalization and by (ii) a  convolutional layer with no activation. Both convolutional layers have kernel size $3$. The output of this procedure is then summed with the residual, which is obtained by processing the input with a convolutional layer with kernel size $1$. 

The encoder part of the network consists of $3$ downsampling blocks, each consisting of $4$ residual blocks having filter sizes $64,128,256$. The output of each downsampling block is followed by average pooling, with pool size $2$ in order to compress the dimension of the spectrograms. The last block of the encoder is followed a self-attention block.

The bottleneck obtained through the encoder is processed by a residual block with $512$ filters and is then processed by the decoder, which is a specular version of the encoder. The only difference lies in the use of transposed convolutions in order to create upsampling layers needed to increase the dimension of the features.

The last downsampling layer of the encoder, the bottleneck and the first upsampling layer of the decoder are followed by self-attention.

\subsection{Deployment} 
The proposed model takes as input spectrograms of a fixed size, therefore audio tracks longer than the ones used for training need to be sliced accordingly. 

The decoder takes as input the conditioning spectrogram $\mathbf{X}$ and the diffusion noise and retrieves an estimate of the latter, which can then be subtracted in order to obtain an estimate of the desired output timbre spectrogram $\hat{\mathbf{Y}}$. The output waveform $y$ can then be obtained by feeding the pre-trained SoundStream model with $\hat{\mathbf{Y}}$.

\section{EXPERIMENTS}

In this section, we describe experiments performed with the aim of demonstrating the capabilities of the proposed DiffTransfer technique both in the single-instrument and multi-instrument application scenarios.

\begin{figure}[t]
    \centering
    \subfloat[]{
    \includegraphics[width=.45\textwidth]{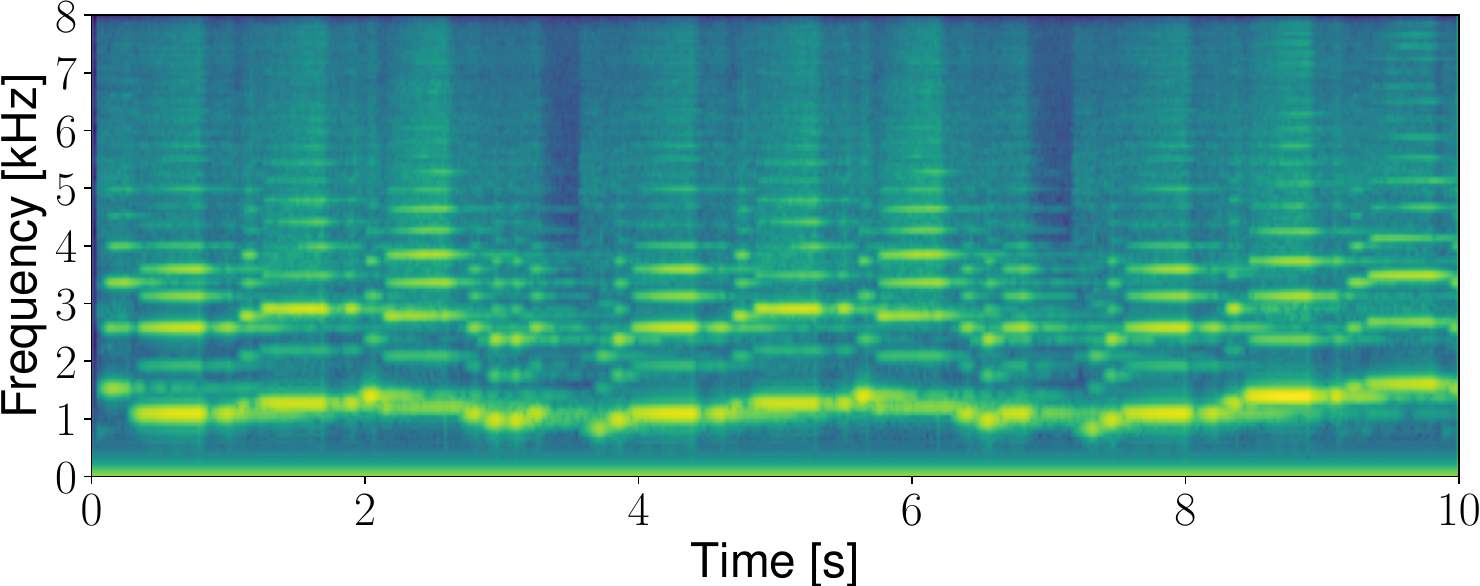}
    \label{subfig:cond}
    }
    
    \centering
    \subfloat[]{
    \includegraphics[width=.45\textwidth]{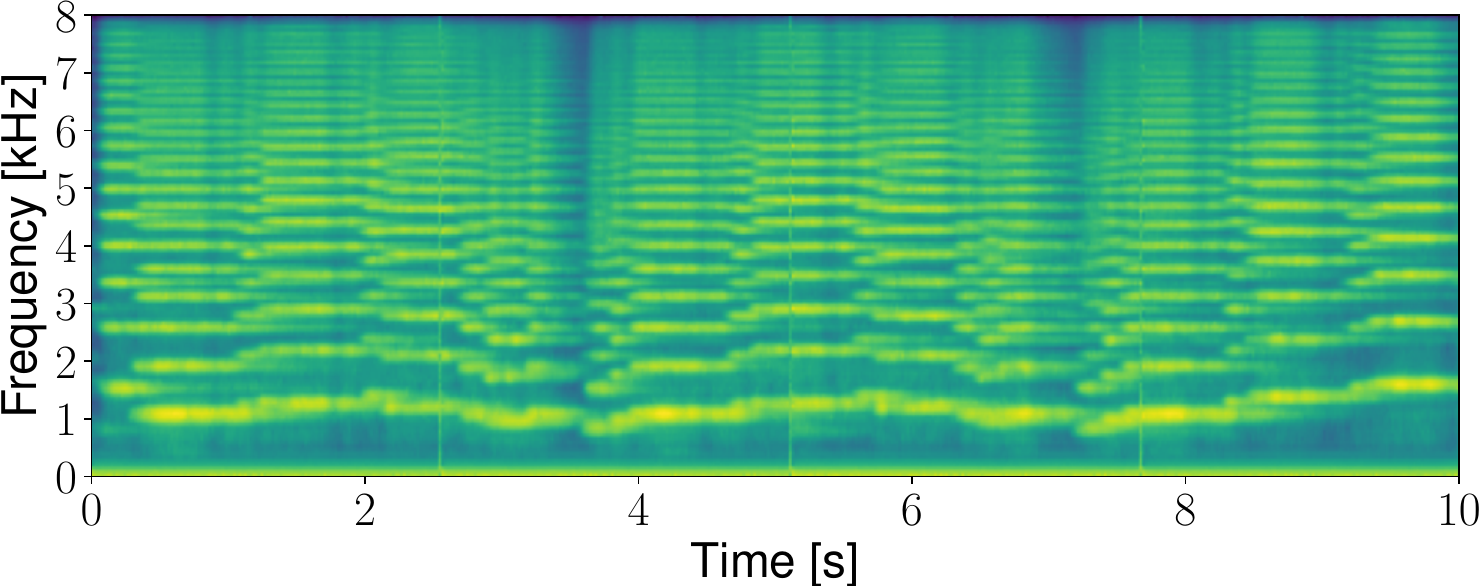}
    \label{subfig:est}
    }

    \centering
    \subfloat[]{
\includegraphics[width=.45\textwidth]{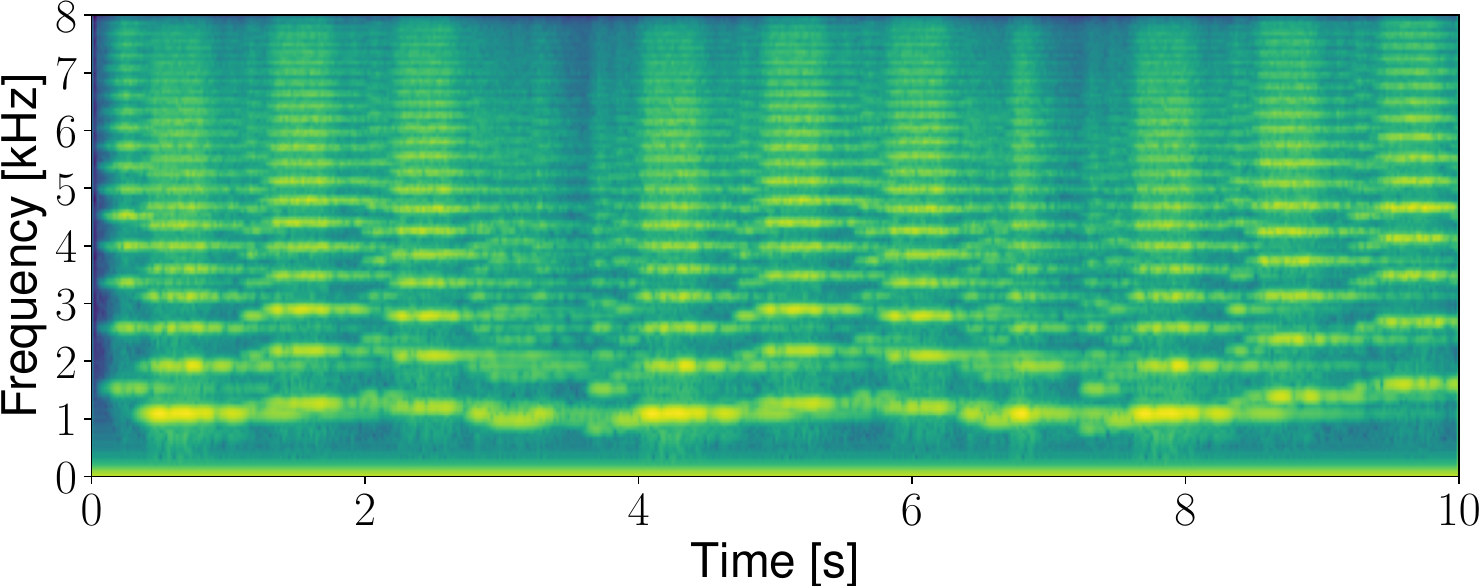}
    \label{subfig:gt}
    }
    \caption{Example of Timbre Conversion log mel Spectrograms using the DiffTransfer architecture, obtained when converting  Clarinet \protect\subref{subfig:cond} to Strings \protect\subref{subfig:est}. The ground truth Strings spectrogram is shown in \protect\subref{subfig:gt}.}
\label{fig:example_specs}
\end{figure}

In Fig.~\ref{fig:example_specs} we show an example of input, generated and ground-truth spectrograms, obtained via the DiffTransfer model when converting from a  Clarinet to Strings.
\subsection{Dataset}
In order to train the model we considered the StarNet dataset~\cite{alinoori_mahshid_2022_6917099}, which contains a set of tracks that are played with two timbre-domains, namely strings-piano and vibraphone-clarinet. The dataset consists of roughly 22 hours of audio. We used the reduced version of the dataset, where tracks are resampled to $16000 ~\mathrm{Hz}$ and converted them to mono. In order to perform the evaluation, we use the same ten tracks considered in~\cite{alinoori2022musicstar}, in order to ease the comparison with their model. 

\subsection{Techniques Under Comparison}
We consider two baselines in order to compare the performances of the proposed DiffTransfer architecture. For what concerns the single-instrument timbre transfer task, we consider the Universal Network~\cite{musictranslation} fine-tuned on the StarNet dataset as done in~\cite{alinoori2022musicstar}. For what concerns the multi-timbre task, we consider the mixture-supervised version of the Music-STAR network proposed in~\cite{alinoori2022musicstar}. We perform three different types of timbre transfer tasks: \textit{single}, where only single instruments are converted, \textit{single/mixed} where the separate conversions of single instruments are mixed in order to create the desired mixture track and \textit{mixture}, where the mixture is directly converted. These nomenclatures are used just to ease the presentation of the results, we would like to point out that, for what concerns the DiffTransfer architecture, no specific changes are required for the various types of applications, except for the choice of 
desired input data.

\subsection{Experiment Setup}
The Universal Network and Music-STAR architectures are trained with the procedure described in~\cite{alinoori2022musicstar}. The DiffTransfer network is trained for $5000$ epochs using a batch size of $16$, with the AdamW optimizer~\cite{loshchilov2018decoupled} with learning rate $2e-5$ and weight decay $1e-4$. The epoch that minimizes the $L1$ noise prediction loss is chosen in order to retain the model used to compute the results. We train a total of six models, performing the following timbre transfer conversions: vibraphone to piano, piano to vibraphone, clarinet to strings, strings to clarinet
vibraphone/clarinet to piano/strings and piano/strings to vibraphone/clarinet.

The network input features are computed by first applying the Short-Time Fourier Transform (STFT) with a Hann window of size $0.020~\mathrm{s}$ and $50 \%$ overlap to normalized audio tracks. Then the log mel spectrogram is computed over $128$ bins corresponding to the range of $0-16000~\mathrm{Hz}$. We do not feed the entire audio tracks as input to the network, instead, during each epoch we extract $128$ frames from the log mel spectrogram, corresponding to $\approx~2~\mathrm{s}$. Each spectrogram slice is normalized between $-1$ and $1$ before being given as input to the network and the output spectrograms are denormalized before being fed to the SoundStream model in order to recover the audio waveform. Since the tracks considered for the test are of length $10~\mathrm{s}$ and the model gets as input a fixed $128$ frames spectrogram we slice the conditioning spectrogram before feeding into the model and we keep the input noise fixed for all slices, in order to ensure consistency in the generation. All spectrogram slices are normalized in the range $[-1, 1]$ and denormalized before being fed to the SoundStream decoder.

\subsection{Objective Evaluation}
We evaluate the model objectively in order to analyze the perceptual similarity and content preservation capabilities of the generated tracks with respect to the ground truth audio.
\begin{table}[t]
    \centering
\begin{tabular}{ |c|c|c|c|  }
 \hline
 \multicolumn{3}{|c|}{\textbf{Objective Evaluation}} \\
 \hline
 \textbf{Method}& \textbf{FAD} $\downarrow$ & \textbf{JD} $\downarrow$\\
 \hline
 Universal Network (single)   & 7.09  & 0.53   \\
 \hline
 DiffTransfer (single)&     2.58 &    0.28\\
 \hline
 \hline
 Universal Network (single/mixed) &10.47  &0.64 \\
 \hline
 DiffTransfer (single/mixed) &4.73 &0.46\\
 \hline
 \hline
 Music-STAR (mixture) & 8.93 &0.57\\
 \hline
 DiffTransfer (mixture) & 4.37 & 0.38\\
 \hline
\end{tabular}
    \caption{Objective Evaluation of the proposed DiffTransfer Method compared to the baselines, in terms of Fr\'{e}chet Audio Distance (FAD) and Jaccard Distance (JD). Results are averaged over all participants and over all the tracks considered for each part of the test.}
    \label{tab:objective_results}
\end{table}

In order to evaluate the perceptual similarity, we compute the Fr\'{e}chet Audio Distance (FAD)~\cite{kilgour2019frechet} using the VGGish embeddings~\cite{hershey2017cnn}, through a PyTorch implementation\footnote{\url{https://pypi.org/project/frechet-audio-distance/}}. FAD is a reference-free metric for music enhancement algorithms, which views the embeddings as a continous multivariate Gaussian and is computed between the real and generated data as
\begin{equation}
    \mathrm{FAD} = ||\mu_r -\mu_g||^2 + \mathrm{tr}(\Sigma_r + \mu_g -2\sqrt{\Sigma_r \Sigma_g}),
\end{equation}
where $(\mu_r, \Sigma_r)$ and $(\mu_g, \Sigma_g)$ are the mean and covariances of the embeddings corresponding to the real and generated data, respectively.
Similarly to~\cite{hawthorne2022multiinstrument}, we compute FAD in order to analyze the perceptual similarity between the generated audios with respect to the ground truth one, corresponding to the original StarNet dataset.

To understand the content-preservation capabilities of the model, following~\cite{cifka2021self}, we compute how the pitch contours of generated ground truth audio tracks are dissimilar, by calculating the mismatch between two sets of pitches $A$ and $B$ through the Jaccard Distance
\begin{equation}
    JD(A,B) = 1 - \frac{|A \cap B|}{|A \cup B|},
\end{equation}
where a lower value corresponds to a lower mismatch and thus to a higher degree of similarity between the generated pitch contours. Pitch contours are computed using a multi-pitch version of the MELODIA~\cite{salamon2012melody} as implemented in the Essentia library~\cite{bogdanov2013essentia}, rounding pitches to the nearest semitone. We report the values obtained by computing the metrics on the test dataset in Table~\ref{tab:objective_results}.

\subsection{Subjective Evaluation}
In order to evaluate subjectively the timbre transfer capabilities, we perform a listening test with 18 human participants. The web page of the test is available at \footnote{\url{https://listening-test-ismir-ttd.000webhostapp.com/}}. The test was split into two parts corresponding to the single and multiple instrument application scenarios, respectively.

During the single instrument part of the test, the users listened to four tracks, corresponding to the four types of conversions performed, namely: clarinet to strings, strings to clarinet, piano to vibraphone, vibraphone to piano. Each example consisted of two conditions, one obtained via the DiffTransfer model and the other through the Universal Network.

In the second part of the test, concerning multiple instrument timbre transfer, a total of four tracks were considered, two for the conversion from vibraphone/strings to piano/strings waveforms and two for the reverse conversion. Each example consisted of four conditions, namely DiffStar (single/mix), Universal Network (single/mix), DiffStar (mixture) and Music-STAR (mixture).

Both the order of conditions and the order of examples in each separate part of the test were randomized.

The participants were asked to rate the conditions in terms of similarity with respect to the reference track on a 5 elements Likert scale where $1$ corresponds to bad and $5$ to excellent.
We report the results obtained through the listening test in Table~\ref{tab:subjective_results}.

\subsection{Discussion}
\begin{table}[t]
    \centering
\begin{tabular}{ |c|c|c|  }
  \hline
 \multicolumn{2}{|c|}{\textbf{Subjective Evaluation}} \\
 \hline
 \textbf{Method}& \textbf{Similarity} \\
 \hline
 \hline
 Universal Network (single)  & 1.82   \\
 \hline
 DiffTransfer (single)&       3.68\\
 \hline
 \hline
 Universal Network (single/mixed)  &1.69 \\
 \hline
 DiffTransfer (single/mixed)&3.78\\
 \hline
 \hline
 Music-STAR (mixture) &2.89\\
 \hline
 DiffTransfer (mixture) &3.80\\
 \hline
\end{tabular}
    \caption{Objective Evaluation of the proposed DiffTransfer Method compared to the baselines, in terms of perceived similarity with respect to the ground truth on a Likert scale from 1 (Bad) to 5 (Excellent). Results are averaged over all test tracks.}
    \label{tab:subjective_results}
\end{table}
By briefly inspecting both the objective and subjective results, reported in Table~\ref{tab:objective_results} and ~\ref{tab:subjective_results}, respectively, it is clear how the proposed DiffTransfer model outperforms the Universal Network and Music-STAR baselines both for what concerns the single and multiple timbre transfer tasks. 

When considering single timbre results, DiffTransfer is able to achieve significantly better performances in terms of FAD, Jaccard Distance and Perceived Similarity, with respect to the Universal network. The gap between the two methods becomes even more evident when considering the single/mixed case, i.e. when single timbre transfer tracks are mixed in order to form the desired mixture audio.

For what concerns the Music-STAR method, the gap with respect to DiffTransfer remains high in terms of FAD, but becomes less noticeable when considering JD and the perceived subjective similarity.

\section{Conclusion}
In this paper, we have presented DiffTransfer a technique for both single- and multi-instrument timbre transfer using Denoising Diffusion Implicit models. The novelty of the proposed approach lies in the fact that in addition to being, to the best of our knowledge, the first application of diffusion models to timbre transfer, it is the first model to be tested in order to perform single and multi-timbre transfer, without varying the architecture depending on which application is chosen.
We compared the proposed model with state-of-the-art Universal Network and Music-STAR baselines through both objective evaluation measures and a listening test, demonstrating the better capabilities of the proposed DiffTransfer approach.

Future works will involve increasing the audio quality of the generated audio, by taking into account the consistency of subsequent generated spectrograms. Furthermore, we plan on modifying the model in order to be able to perform unpaired timbre transfer, which greatly eases the dataset requirements and applicability of the technique. 
\bibliography{ISMIRtemplate}

\begin{thebibliography}{10}
\providecommand{\url}[1]{#1}
\csname url@samestyle\endcsname
\providecommand{\newblock}{\relax}
\providecommand{\bibinfo}[2]{#2}
\providecommand{\BIBentrySTDinterwordspacing}{\spaceskip=0pt\relax}
\providecommand{\BIBentryALTinterwordstretchfactor}{4}
\providecommand{\BIBentryALTinterwordspacing}{\spaceskip=\fontdimen2\font plus
\BIBentryALTinterwordstretchfactor\fontdimen3\font minus
  \fontdimen4\font\relax}
\providecommand{\BIBforeignlanguage}[2]{{%
\expandafter\ifx\csname l@#1\endcsname\relax
\typeout{** WARNING: IEEEtran.bst: No hyphenation pattern has been}%
\typeout{** loaded for the language `#1'. Using the pattern for}%
\typeout{** the default language instead.}%
\else
\language=\csname l@#1\endcsname
\fi
#2}}
\providecommand{\BIBdecl}{\relax}
\BIBdecl

\bibitem{colonel2020conditioning}
J.~T. Colonel and S.~Keene, ``Conditioning autoencoder latent spaces for
  real-time timbre interpolation and synthesis,'' in \emph{2020 International
  Joint Conference on Neural Networks (IJCNN)}.\hskip 1em plus 0.5em minus
  0.4em\relax IEEE, 2020, pp. 1--7.

\bibitem{song2021denoising}
J.~Song, C.~Meng, and S.~Ermon, ``Denoising diffusion implicit models,'' in
  \emph{International Conference on Learning Representations}, 2021.

\bibitem{saharia2022palette}
C.~Saharia, W.~Chan, H.~Chang, C.~Lee, J.~Ho, T.~Salimans, D.~Fleet, and
  M.~Norouzi, ``Palette: Image-to-image diffusion models,'' in \emph{ACM
  SIGGRAPH 2022 Conference Proceedings}, 2022, pp. 1--10.

\bibitem{huang2018timbretron}
S.~Huang, Q.~Li, C.~Anil, X.~Bao, S.~Oore, and R.~B. Grosse, ``Timbretron: A
  wavenet(cycle{GAN}({CQT}(audio))) pipeline for musical timbre transfer,'' in
  \emph{International Conference on Learning Representations}, 2019.

\bibitem{zhu2017unpaired}
J.-Y. Zhu, T.~Park, P.~Isola, and A.~A. Efros, ``Unpaired image-to-image
  translation using cycle-consistent adversarial networks,'' in \emph{Proc. of
  the IEEE international conference on computer vision}, 2017, pp. 2223--2232.

\bibitem{oord2016wavenet}
A.~v.~d. Oord, S.~Dieleman, H.~Zen, K.~Simonyan, O.~Vinyals, A.~Graves,
  N.~Kalchbrenner, A.~Senior, and K.~Kavukcuoglu, ``Wavenet: A generative model
  for raw audio,'' \emph{arXiv preprint arXiv:1609.03499}, 2016.

\bibitem{jain2020att}
D.~K. Jain, A.~Kumar, L.~Cai, S.~Singhal, and V.~Kumar, ``Att: Attention-based
  timbre transfer,'' in \emph{2020 International Joint Conference on Neural
  Networks (IJCNN)}.\hskip 1em plus 0.5em minus 0.4em\relax IEEE, 2020, pp.
  1--6.

\bibitem{kumar2019melgan}
K.~Kumar, R.~Kumar, T.~De~Boissiere, L.~Gestin, W.~Z. Teoh, J.~Sotelo,
  A.~de~Br{\'e}bisson, Y.~Bengio, and A.~C. Courville, ``Melgan: Generative
  adversarial networks for conditional waveform synthesis,'' \emph{Advances in
  neural information processing systems}, vol.~32, 2019.

\bibitem{luo2019learning}
Y.-J. Luo, K.~Agres, and D.~Herremans, ``Learning disentangled representations
  of timbre and pitch for musical instrument sounds using gaussian mixture
  variational autoencoders,'' in \emph{20th International Society for Music
  Information Retrieval (ISMIR2019)}, 2019.

\bibitem{Engel2020DDSP}
J.~Engel, L.~H. Hantrakul, C.~Gu, and A.~Roberts, ``Ddsp: Differentiable
  digital signal processing,'' in \emph{International Conference on Learning
  Representations}, 2020.

\bibitem{michelashvili2020timbre}
M.~Michelashvili and L.~Wolf, ``Hierarchical timbre-painting and articulation
  generation,'' in \emph{21st International Society for Music Information
  Retrieval (ISMIR2020)}, 2020.

\bibitem{musictranslation}
A.~P. Noam~Mor, Lior~Wold and Y.~Taigman, ``A universal music translation
  network,'' in \emph{International Conference on Learning Representations
  (ICLR)}, 2019.

\bibitem{alinoori2022musicstar}
M.~Alinoori and V.~Tzerpos, ``Music-star: a style translation system for
  audio-based re-instrumentation,'' in \emph{21st International Society for
  Music Information Retrieval (ISMIR2022)}, 2022.

\bibitem{engel2017neural}
J.~Engel, C.~Resnick, A.~Roberts, S.~Dieleman, M.~Norouzi, D.~Eck, and
  K.~Simonyan, ``Neural audio synthesis of musical notes with wavenet
  autoencoders,'' in \emph{International Conference on Machine Learning}.\hskip
  1em plus 0.5em minus 0.4em\relax PMLR, 2017, pp. 1068--1077.

\bibitem{williams1989learning}
R.~J. Williams and D.~Zipser, ``A learning algorithm for continually running
  fully recurrent neural networks,'' \emph{Neural computation}, vol.~1, no.~2,
  pp. 270--280, 1989.

\bibitem{ho2020denoising}
J.~Ho, A.~Jain, and P.~Abbeel, ``Denoising diffusion probabilistic models,''
  \emph{Advances in Neural Information Processing Systems}, vol.~33, pp.
  6840--6851, 2020.

\bibitem{goodfellow2020generative}
I.~Goodfellow, J.~Pouget-Abadie, M.~Mirza, B.~Xu, D.~Warde-Farley, S.~Ozair,
  A.~Courville, and Y.~Bengio, ``Generative adversarial networks,''
  \emph{Communications of the ACM}, vol.~63, no.~11, pp. 139--144, 2020.

\bibitem{roberts2017hierarchical}
A.~Roberts, J.~Engel, and D.~Eck, ``Hierarchical variational autoencoders for
  music,'' in \emph{NIPS Workshop on Machine Learning for Creativity and
  Design}, vol.~3, 2017.

\bibitem{saharia2022photorealistic}
C.~Saharia, W.~Chan, S.~Saxena, L.~Li, J.~Whang, E.~Denton, S.~K.~S.
  Ghasemipour, R.~Gontijo-Lopes, B.~K. Ayan, T.~Salimans, J.~Ho, D.~J. Fleet,
  and M.~Norouzi, ``Photorealistic text-to-image diffusion models with deep
  language understanding,'' in \emph{Advances in Neural Information Processing
  Systems}, A.~H. Oh, A.~Agarwal, D.~Belgrave, and K.~Cho, Eds., 2022.

\bibitem{hawthorne2022multiinstrument}
C.~Hawthorne, I.~Simon, A.~Roberts, N.~Zeghidour, J.~Gardner, E.~Manilow, and
  J.~Engel, ``Multi-instrument music synthesis with spectrogram diffusion,'' in
  \emph{Ismir 2022 Hybrid Conference}, 2022.

\bibitem{yang2022diffsound}
D.~Yang, J.~Yu, H.~Wang, W.~Wang, C.~Weng, Y.~Zou, and D.~Yu, ``Diffsound:
  Discrete diffusion model for text-to-sound generation,'' 2022.

\bibitem{mittal2021symbolicdiffusion}
G.~Mittal, J.~Engel, C.~Hawthorne, and I.~Simon, ``Symbolic music generation
  with diffusion models,'' in \emph{Proc. of the 22nd International Society for
  Music Information Retrieval Conference}, 2021.

\bibitem{kong2021diffwave}
Z.~Kong, W.~Ping, J.~Huang, K.~Zhao, and B.~Catanzaro, ``Diffwave: A versatile
  diffusion model for audio synthesis,'' in \emph{International Conference on
  Learning Representations}, 2021.

\bibitem{PlajaRoglans2022}
G.~Plaja-Roglans, M.~Miron, and X.~Serra, ``A diffusion-inspired training
  strategy for singing voice extraction in the waveform domain,'' in
  \emph{International Society for Music Information Retrieval (ISMIR)
  Conference}, 2022.

\bibitem{zhang2023towards}
G.~Zhang, J.~Ji, Y.~Zhang, M.~Yu, T.~Jaakkola, and S.~Chang, ``Towards coherent
  image inpainting using denoising diffusion implicit models,'' \emph{arXiv
  preprint arXiv:2304.03322}, 2023.

\bibitem{zeghidour2021soundstream}
N.~Zeghidour, A.~Luebs, A.~Omran, J.~Skoglund, and M.~Tagliasacchi,
  ``Soundstream: An end-to-end neural audio codec,'' \emph{IEEE/ACM
  Transactions on Audio, Speech, and Language Processing}, vol.~30, pp.
  495--507, 2021.

\bibitem{campbell2022continuous}
A.~Campbell, J.~Benton, V.~De~Bortoli, T.~Rainforth, G.~Deligiannidis, and
  A.~Doucet, ``A continuous time framework for discrete denoising models,''
  \emph{Advances in Neural Information Processing Systems}, vol.~35, pp.
  28\,266--28\,279, 2022.

\bibitem{song2021scorebased}
Y.~Song, J.~Sohl-Dickstein, D.~P. Kingma, A.~Kumar, S.~Ermon, and B.~Poole,
  ``Score-based generative modeling through stochastic differential
  equations,'' in \emph{International Conference on Learning Representations},
  2021.

\bibitem{rouard2021crash}
S.~Rouard and G.~Hadjeres, ``Crash: raw audio score-based generative modeling
  for controllable high-resolution drum sound synthesis,'' in \emph{22nd
  International Society for Music Information Retrieval (ISMIR2021)}, 2021.

\bibitem{nichol2021improved}
A.~Q. Nichol and P.~Dhariwal, ``Improved denoising diffusion probabilistic
  models,'' in \emph{International Conference on Machine Learning}.\hskip 1em
  plus 0.5em minus 0.4em\relax PMLR, 2021, pp. 8162--8171.

\bibitem{alinoori_mahshid_2022_6917099}
\BIBentryALTinterwordspacing
M.~Alinoori and V.~Tzerpos, ``Starnet,'' Aug. 2022. [Online]. Available:
  \url{https://doi.org/10.5281/zenodo.6917099}
\BIBentrySTDinterwordspacing

\bibitem{loshchilov2018decoupled}
I.~Loshchilov and F.~Hutter, ``Decoupled weight decay regularization,'' in
  \emph{International Conference on Learning Representations}, 2019.

\bibitem{kilgour2019frechet}
K.~Kilgour, M.~Zuluaga, D.~Roblek, and M.~Sharifi, ``Fr{\'e}chet audio
  distance: A reference-free metric for evaluating music enhancement
  algorithms.'' in \emph{INTERSPEECH}, 2019, pp. 2350--2354.

\bibitem{hershey2017cnn}
S.~Hershey, S.~Chaudhuri, D.~P. Ellis, J.~F. Gemmeke, A.~Jansen, R.~C. Moore,
  M.~Plakal, D.~Platt, R.~A. Saurous, B.~Seybold \emph{et~al.}, ``Cnn
  architectures for large-scale audio classification,'' in \emph{2017 ieee
  international conference on acoustics, speech and signal processing
  (icassp)}.\hskip 1em plus 0.5em minus 0.4em\relax IEEE, 2017, pp. 131--135.

\bibitem{cifka2021self}
O.~C{\'\i}fka, A.~Ozerov, U.~{\c{S}}im{\c{s}}ekli, and G.~Richard,
  ``Self-supervised vq-vae for one-shot music style transfer,'' in \emph{ICASSP
  2021-2021 IEEE International Conference on Acoustics, Speech and Signal
  Processing (ICASSP)}.\hskip 1em plus 0.5em minus 0.4em\relax IEEE, 2021, pp.
  96--100.

\bibitem{salamon2012melody}
J.~Salamon and E.~G{\'o}mez, ``Melody extraction from polyphonic music signals
  using pitch contour characteristics,'' \emph{IEEE transactions on audio,
  speech, and language processing}, vol.~20, no.~6, pp. 1759--1770, 2012.

\bibitem{bogdanov2013essentia}
D.~Bogdanov, N.~Wack, E.~G{\'o}mez~Guti{\'e}rrez, S.~Gulati, H.~Boyer,
  O.~Mayor, G.~Roma~Trepat, J.~Salamon, J.~R. Zapata~Gonz{\'a}lez, X.~Serra
  \emph{et~al.}, ``Essentia: An audio analysis library for music information
  retrieval,'' in \emph{Britto A, Gouyon F, Dixon S, editors. 14th Conference
  of the International Society for Music Information Retrieval (ISMIR); 2013
  Nov 4-8; Curitiba, Brazil.[place unknown]: ISMIR; 2013. p. 493-8.}\hskip 1em
  plus 0.5em minus 0.4em\relax International Society for Music Information
  Retrieval (ISMIR), 2013.

\end{thebibliography}
\end{document}